\documentclass[
11pt,%
tightenlines,%
twoside,%
onecolumn,%
nofloats,%
nobibnotes,%
nofootinbib,%
superscriptaddress,%
noshowpacs,%
centertags]%
{revtex4}

\usepackage{ljm}

\begin{document}

\titlerunning{H-theorem for systems}

\authorrunning{A.~V.\ Lebedev and G.~B.\ Lesovik}

\title {H-theorem for Systems with an Interaction Invariant Distribution Function}

\author{\firstname {A.~V.}~\surname {Lebedev} and \firstname{G.~B.}~\surname{Lesovik}}
\email[E-mail:]{alebedev79@bk.ru}
\affiliation {Moscow Institute of Physics and Technology, Institutskii per.\ 9, Dolgoprudny, 141700, Moscow District, Russia}

\firstcollaboration{Submitted by S. A. Grigoryan}

\received{May 8, 2019;  revised May 13, 2019; accepted May 18, 2019}

\begin {abstract}{H-theorem gives necessary conditions for a system to evolve in time  with a non-diminishing
entropy. In a quantum case the role of H-theorem plays the unitality criteria of a quantum channel transformation describing the evolution of the
system's density matrix under the presence of the interaction with an environment. Here, we show that if diagonal elements of the system's density matrix are
robust to the presence of interaction the corresponding quantum channel is unital.}
\end{abstract}



\maketitle

\section{Introduction}

The second law of thermodynamics poses a constrain on the time
direction of natural thermodynamic processes. Its classical
formulation states that the entropy of an isolated system can not
decrease over time. The celebrated H-theorem of Ludwig Boltzmann
\cite{boltzmann1,boltzmann2} guarantees the non-diminishing entropy
increase of a classical system those dynamics is governed by the
kinetic equation of a specific form. Boltzmann's kinetic equations
relies on the molecular chaos hypothesis which assumes the chaotic
uncorrelated initial state of colliding particles
\cite{lebowitz:1999}. Therefore, the Boltzmann's H-theorem can be
viewed as a sufficient constraint on dynamics of classical particles
to develop with a non-diminishing entropy.

In a quantum realm the entropy of an isolated quantum system stays
constant and the classical formulation of the second law is trivial.
Therefore, to make the meaningful formulation of the second law in
quantum case one needs to consider the entropy dynamics of an open
quantum system. Here, the quantum master equation \cite{breuer}
describing the evolution of the system's density matrix $\hat\rho$
is an analog of the Boltzmann's kinetic equation for a classical
case. For a quantum system interacting with a memoryless environment
(or reservoir) the evolution of the system's density matrix can be
described by Marcovian (or equivalently Lindblad) master equation
\cite{lindblad:1976}. In this situation one can prove the
non-negativity of the \textit{entropy production rate}
$\sigma(\hat\rho)$ that is the amount of entropy produced per unit
time by the system \cite{lindblad:1975}. In non-equilibrium
thermodynamics $\sigma(\hat\rho)$ obeys a balance equation $\sigma =
\frac{d}{dt} S\bigl(\hat\rho_t\bigr) + J$, where $S(\hat\rho) =
-k_B\mbox{tr}\{ \hat\rho \ln\hat\rho \}$ is the von Neumann entropy
and $J$ is the entropy flux that is an amount of entropy which is
exchanged between the open system and reservoir per unit time. For a
thermal equilibrium reservoir at a temperature $T$ the entropy flux
$J = - \frac1{T}\frac{d Q}{dt},$ where $Q$ id the heat dissipated by
the system into the environment. For an energy isolated open system
$J = 0$ and the inequality $\sigma(\hat\rho(t)) \geq 0$ expresses
the second law of thermodynamics.

The quantum information theory \cite{nielsen,holevo:2012} (QIT)
suggests  a more general approach to the dynamics of open quantum
systems which goes far beyond the Marcovian approximation. Within
the QIT framework the evolution of the system's density matrix is
described via the so called \textit{quantum channel}: $\rho_0 \to
\Phi(\rho_0) \equiv \hat\rho_t$ that is a completely positive,
trace-preserving linear map of a density matrix $\hat\rho$. Given
the uncorrelated initial product state $\hat\rho_0 \otimes
\hat\pi_0$ of the system's density matrix $\hat\rho_0$ and its
environment $\hat\pi_0$, the explicit form of the quantum channel
can be found by tracing out the environmental degrees of freedom in
a time-evolved state of a grand system (a given quantum system plus
its environment),
$
      \Phi(\rho_0) = \mbox{Tr}_\mathrm{env} \{ \hat{U}_t (\hat\rho_0 \otimes \hat\pi_0)  \hat{U}_t^\dagger\bigr\},
      \nonumber
$
where a joint unitary evolution operator $\hat {U}_t =
\exp(-it\hat{H}/\hbar)$ is generated by  the Hamiltonian $\hat{H} =
\hat{H}_\mathrm{S} + \hat{H}_\mathrm{R} + \hat{H}_\mathrm{int}$
comprised of the free Hamiltonians of the quantum system
$\hat{H}_\mathrm{S}$ and the reservoir $\hat{H}_\mathrm{R}$ and the
interaction term $\hat{H}_\mathrm{int}$. In this context the
Lindbald dynamics can be described within the so called
\textit{collision model} \cite{ziman}, where a quantum system
interacts locally in time and only once with different uncorrelated
environmental degrees of freedom or sub-environments. The resulting
quantum channel then possesses a \textit{divisibility} property:
$\Phi = \Phi_N \circ \dots \circ \Phi_1$, where $\Phi_i$ is a
quantum channel corresponding to the interaction with an $i$th
sub-environment followed by a free evolution of a quantum system.

However, in the presence of a correlated environment \cite{rubar}
and/or finite  time reservoir memory effects \cite{ciccarello} the
quantum channel is not divisible and hence cannot be described by
the Lindblad master equation. This may result in a non-monotonic
entropy dynamics and the negative entropy gain in general. One of
the remarkable results of the QIT is the lower bound for the entropy
gain of an arbitrary quantum channel \cite{holevo:2010},
\begin{equation}
      S\bigl( \Phi(\hat\rho) \bigr) - S(\hat\rho) = -k_B \mbox{Tr}\{ \Phi(\hat\rho) \ln\Phi(\hat{1})\},
      \label{eq:entropy_gain}
\end{equation}
where $\hat{1}$ is the identity operator. In \cite{Amosov:2015} this
result was further generalized for a tensor product of a dephasing
channel and an arbitrary quantum channel. It follows from the right
hand side of (\ref{eq:entropy_gain}) that for a wide class the so
called \textit{unital} quantum channels defined by the relation
$\Phi(\hat{1}) = \hat{1}$ the entropy gain is non-negative.

In \cite{lesovik:2016} the unitality constraint has been considered
as a quantum analog of the Boltzmann's H-theorem and the unitality
criteria was formulated in terms of the joint system-reservoir
unitary evolution operators. Consider a decomposition of the unitary
evolution operator $\hat{U}_t$ of the grand system in interaction
representation $\hat{U}_t = \hat{U}_\mathrm{S} \hat{U}_\mathrm{R}
\hat{U}_\mathrm{int}$, where $\hat{U}_\mathrm{S} =
\exp(-it\hat{H}_\mathrm{S}/\hbar)$ and $\hat{U}_\mathrm{R} =
\exp(-it\hat{H}_\mathrm{R}/\hbar)$ are the free evolution operators
of the two subparts of the grand system and $\hat{U}_\mathrm{int} =
T\exp(-i\int^t dt^\prime \,\hat{H}_\mathrm{int}(t^\prime)/\hbar)$ is
the joint evolution operator. Then the operator $\hat{U}_t$ can be
represented in the factorised form, $
      \hat{U}_t = \sum_{ki} |\tilde\psi_k\rangle \langle \psi_i|\, \hat{U}_\mathrm{R} \hat{V}_{ki},
      $
where
\begin{equation}
      \hat{V}_{ki} \equiv \langle \psi_k| \hat{U}_\mathrm{int} |\psi_i\rangle
      \label{eq:vki}
\end{equation}
is a set of operators acting in the reservoir Hilbert space,
$|\psi_i\rangle$ is a complete orthonormal  set of states in the
system's Hilbert space and $|\tilde\psi_k\rangle \equiv
\hat{U}_\mathrm{S}|\psi_k\rangle$. Then the unitality condition can
be expressed as
\begin{equation}
      \bigl[\Phi(\hat{1})\bigr]_{kk^\prime} - \delta_{kk^\prime} = \sum_{i}\Bigl\langle\bigl[ \hat{V}_{k^\prime i}^\dagger, \hat{V}_{ki} \bigr] \Bigr\rangle,
      \label{eq:htheorem}
\end{equation}
where $\bigl[\Phi(\hat{1})\bigr]_{kk^\prime} = \langle \tilde\psi_k|
\Phi(\hat{1})|\tilde\psi_{k^\prime}\rangle$. The necessary and
sufficient requirement for the channel $\Phi$ to be unital then can
be found in the vanishing of the right hand side of
(\ref{eq:htheorem}).  This unitality criteria was found to be an
effective tool in describing the entropy production in exemplary
physical phenomena like electron inelastic scattering,
electron-phonon interaction and so on, see \cite{lesovik:2016}.
Moreover, for an energetically isolated quantum system this criteria
can identify a non-unital channels preserving the energy of the
system and which can lower its entropy without heat exchange with
the reservoir. Such a situation corresponds to an action of a
quantum Maxwell demon \cite{lloyd} and was demonstrated in the
system of interacting qubits \cite{kirsanov:2018}. In this article
we prove a new unitality criteria which is expressed in terms  of
invariance of the diagonal density matrix elements i.e. classical
distribution function of the system to the presence of the
interaction with an environment. Namely, we prove the following

\textbf{Theorem.} \textit{Consider a quantum system endowed with a
finite  dimensional Hilbert space initially uncorrelated and
disentangled from a reservoir. If there is a basis where for all
initial states of the system the diagonal elements of its density
matrix with and without interaction with the reservoir coincide at
the end of the evolution then the system evolves under a unital
quantum channel.}

First, we prove the following

\textbf{Lemma.} \textit{Let all the requirements of the Theorem hold
true and $|\tilde\psi_k\rangle$ is the specific basis, where}
\begin{equation}
    \forall \hat\rho, k \qquad \langle \tilde\psi_k| \Phi(\hat\rho)|\tilde\psi_k\rangle = \langle \tilde\psi_k| \Phi_0(\hat\rho)|\tilde\psi_k\rangle,
    \label{eq:req}
\end{equation}
\textit{$\Phi(\hat\rho)$ and $\Phi_0(\hat\rho)$ are the quantum
channels describing the evolution of the quantum system with and
without interaction with the reservoir initially prepared in a state
$\hat\pi_0$. Let $\hat{U}_\mathrm{S}$ and $\hat{U}_\mathrm{R}$ are
the free evolution operators for the system and the reservoir
respectively, and $\hat{U}_\mathrm{int}$ is the interacting part of
the global evolution operator $\hat{U} = \hat{U}_\mathrm{S}
\hat{U}_\mathrm{R} \hat{U}_\mathrm{int}$. Then for any eigenstate
$|n\rangle$ of the initial density matrix of the reservoir with
$\langle n| \hat\pi_0 |n\rangle > 0$}
\begin{equation}
      \hat{U}_\mathrm{int}\bigl( |\psi_k\rangle \otimes |n\rangle \bigr) = |\psi_k\rangle \otimes |\varphi_{n,k}\rangle \qquad \forall k,
      \nonumber
\end{equation}
\textit{where $|\psi_k\rangle = \hat{U}_\mathrm{S}^\dagger |\tilde\psi_k\rangle$ and $|\varphi_{n,k}\rangle$ is a final normalized state of the reservoir.}

\textit{Proof.} Let the initial state of the grand system comprising
the system and the reservoir has the form
$
      \hat{R}_0 = \sum_{ii^\prime} \rho_{ii^\prime}\, |\psi_i\rangle \langle \psi_{i^\prime}| \otimes \hat\pi_0,
      $
where $\rho_{ii^\prime} = \langle \psi_i| \hat\rho |
\psi_{i^\prime}\rangle$  are the system's density matrix elements.
Then under the unitary evolution $\hat{U}$ the resulting state of
the grand system is
\begin{equation}
      \hat{R} = \sum_{ii^\prime} \rho_{ii^\prime} \sum_{kk^\prime} |\tilde\psi_k\rangle \hat{U}_\mathrm{R} \hat{V}_{ki} \hat\pi_0 \hat{V}_{k^\prime i^\prime}^\dagger \hat{U}_\mathrm{R}^\dagger \langle \tilde\psi_{k^\prime}|,
      \nonumber
\end{equation}
where a set of reservoir's operators $\hat{V}_{ki}$ is defined in
(\ref{eq:vki}).  Tracing out the reservoir degrees of freedom one
gets the quantum channel $\Phi(\hat\rho)$ describing the evolution
in the interacting case
\begin{equation}
      \Phi(\hat\rho) = \sum_{kk^\prime} |\tilde\psi_k\rangle \langle \tilde\psi_{k^\prime}| \sum_{ii^\prime} \rho_{ii^\prime\, }\mbox{Tr}\{ \hat\pi_0 \hat{V}_{k^\prime i^\prime}^\dagger \hat{V}_{ki} \},
      \nonumber
\end{equation}
while in the non-interacting situation one has
$
      \Phi_0(\hat\rho) = \sum_{kk^\prime} |\tilde\psi_k\rangle \langle \tilde\psi_{k^\prime}|\, \rho_{kk^\prime}.
     $
Then the requirement given by (\ref{eq:req}) can be expressed in the
form
$
      \forall k,\hat\rho: \quad \sum_{ij} \rho_{ij} \, \mbox{Tr}\{ \hat\pi_0 \hat{V}_{k j}^\dagger \hat{V}_{ki} \} = \rho_{kk}.
      $
Let us introduce a set of matrixes $H_k$: $ H_{k,ij} \equiv
\mbox{Tr}\{ \hat\pi_0 \hat{V}_{kj}^\dagger \hat{V}_{ki}\}$ and
rewrite (\ref{eq:req}) as
\begin{equation}
      \forall k,\hat\rho: \quad \mbox{Tr}\{ \boldsymbol{\rho} H_k\} = \boldsymbol{\rho}_{kk},
      \label{eq:req2}
\end{equation}
where $\boldsymbol{\rho}$ is the system's density matrix in the
specific representation $|\psi_k\rangle$. The matrixes $H_k$ are
clearly  Hermitian and hence for each $k$ can be diagonalized in a
specific basis
$
      H_{k,ij} = \sum_\alpha h_\alpha^{(k)} \, \xi_{\alpha,i}^{(k)} \, \xi_{\alpha,j}^{(k)*},
      $
where all $h_\alpha^{(k)}$ are real and $\vec\xi_\alpha^{(k)}$ form
a complete and orthonormal  set of eigenvectors. According to the
requirements of the Theorem the Eq. (\ref{eq:req2}) is satisfied for
all $\boldsymbol{\rho}$ and hence for the density matrixes diagonal
in $\vec\xi_\alpha^{(k)}$ representation
%
      $\rho_{ij} = \sum_\alpha p_\alpha \, \xi_{\alpha,i}^{(k)} \,
      \xi_{\alpha,j}^{(k)*}.$
%
Then (\ref{eq:req2}) can be rewritten as
\begin{equation}
      \sum_\alpha h_\alpha^{(k)} p_\alpha = \sum_\alpha p_\alpha \bigl|\xi_{\alpha,k}^{(k)}\bigr|^2,\quad \forall p_\alpha \geq 0, \sum_\alpha p_\alpha = 1.
      \nonumber
\end{equation}
Therefore, $h_\alpha^{(k)} = |\xi_{\alpha,k}^{(k)}|^2$ are
non-negative numbers and as follows from  completeness of the set
$\vec\xi_\alpha^{(k)}$ the matrix $H_k$ has a unit trace:
$\sum_\alpha h_\alpha^{(k)} = \sum_\alpha |\xi_{\alpha,k}^{(k)}|^2 =
1$. Choosing a pure initial state with $\rho_{ij} = 0$  for all
$i,j\neq k$ and $\rho_{kk} = 1$ one gets from the (\ref{eq:req2})
\begin{equation}
      H_{k,kk} = 1, \quad H_{k,ii} = 0 \quad \forall i\neq k.
      \label{eq:Hk}
\end{equation}
Consider a basis in the reservoir Hilbert space diagonalizing the
initial density  matrix $\hat\pi_0$: $\hat\pi_0 = \sum_n \pi_n
|n\rangle \langle n|$. The reservoir operators $\hat{V}_{ki}$ can be
represented as $\hat{V}_{ki}  = \sum_{nm} |n\rangle \langle m|\,
v_{nm}^{(ki)}$ and then the diagonal elements $H_{k,ii} = \sum_n
\pi_n \sum_m |v_{nm}^{(ki)}|^2\geq 0$. Hence, from the Eq.
(\ref{eq:Hk}) it follows $v_{nm}^{(ki)} = 0$ or equivalently
$\hat{V}_{ki}|n\rangle = 0$ for all $i\neq k$. The latter proves the
Lemma:
%
$$
      \forall k:\quad \hat{U}_\mathrm{int} |\psi_k\rangle \otimes |n\rangle
      = \sum_i |\psi_i\rangle \otimes \hat{V}_{ik}|n\rangle = |\psi_k\rangle \otimes |\varphi_{n,k}\rangle,
$$
where the final state of the reservoir $|\varphi_{n,k}\rangle \equiv \hat{V}_{kk}|n\rangle$ is normalized due to $H_{k,kk} = 1$.\qed

\textit{Proof of the Theorem.} Making use the results and
definitions of the above Lemma the  density matrix of the grand
system after interaction with the environment has the form
$$
\hat{R} = \sum_n \pi_n \sum_{kk^\prime} \rho_{kk^\prime}
\hat{U}_\mathrm{S} \hat{U}_\mathrm{R} \hat{U}_\mathrm{int}
|\psi_k\rangle |n\rangle \langle n| \langle \psi_{k^\prime}|
\hat{U}_\mathrm{int}^\dagger \hat{U}_\mathrm{R}^\dagger
\hat{U}_\mathrm{S}^\dagger = \sum_n \pi_n \sum_{kk^\prime}
\rho_{kk^\prime} |\tilde\psi_k\rangle \langle \tilde\psi_{k^\prime}|
\otimes \hat{U}_\mathrm{R}|\varphi_{n,k}\rangle \langle
\varphi_{n,k^\prime}|\hat{U}_\mathrm{R}^\dagger.
$$
Therefore, the induced quantum channel \textit{in the specific basis
set $|\tilde\psi_k\rangle$} has the form
\begin{equation}
      \Phi(\hat\rho) = \sum_n \pi_n \sum_{kk^\prime} \rho_{kk^\prime} |\tilde\psi_k\rangle \langle\tilde\psi_{k^\prime}| \, \langle \varphi_{n,k^\prime}|\varphi_{n,k}\rangle.
      \nonumber
\end{equation}
Introducing a set of Gramian matrixes $
      [\gamma_n]_{kk^\prime} \equiv \langle \varphi_{n,k^\prime}| \varphi_{n,k}\rangle,
      $
the quantum channel $\Phi$ can be rewritten as a convex combination
$
      \Phi(\hat\rho) = \hat{U}_\mathrm{S} \Bigl( \sum_n \pi_n \Phi_n(\hat\rho) \Bigr) \hat{U}_\mathrm{S}^\dagger,
     $
where each channel $\Phi_n$ corresponds to the dephasing channel in the $|\psi_k\rangle = \hat{U}_\mathrm{S}^\dagger |\tilde\psi_k\rangle$ representation:
$
      \Phi_n(\hat\rho) = \sum_{kk^\prime} |\psi_k\rangle \langle \psi_{k^\prime}|\, \rho_{kk^\prime} \bigl[\gamma_n\bigr]_{kk^\prime}.
      $
The dephasing quantum channel is known to be unital
\cite{nielsen,holevo:2012} and hence the convex linear combination
of the dephasing channels is unital as well. This proves the
statement of the Theorem.\qed

We thank G.~M. Graf from ETH Zurich for comments and discussions.
The research was supported by the Government of the Russian
Federation (Agreement 05.Y09.21.0018), by the RFBR Grants No. 17-
02-00396A and 18-02-00642A, Foundation for the Advancement of
Theoretical Physics \textquotedblleft BASIS\textquotedblright, the
Ministry of Education and Science of the Russian Federation
16.7162.2017/8.9 (A.V.L.).

\end{document}